\documentclass[aps,preprint]{revtex4}

\usepackage{amsmath}
\usepackage{booktabs}
\usepackage{float}
\usepackage{graphicx}
\usepackage{soul}

\begin{document}


\title{Comment on ``Improved mutual information measure for clustering, classification, and community detection''}


\author{Zhong-Yuan Zhang}
\email[]{zhyuanzh@gmail.com}
\affiliation{School of Statistics and Mathematics, Central University of Finance and Economics, P.R.China}


\date{\today}

\begin{abstract}
A recent article proposed reduced mutual information for evaluation of clustering, classification and community detection. The motivation is that the standard normalized mutual information (NMI) may give counter-intuitive answers under certain conditions and particularly when the number of clusters differs between the two divisions under consideration. The motivation makes sense. However, the examples given in the article are not accurate, and this comment discusses why. In addition, this comment also empirically demonstrates that the reduced mutual information cannot handle the difficulties of NMI and even brings more. The necessity of Kappa is also empirically validated in this comment.
\end{abstract}


\maketitle

Normalized mutual information (NMI) \cite{danon2005comparing} is widely used for evaluation of clustering and community detection. In a recent article, M. E. J. Newman, George T. Cantwell and Jean-Gabriel Young [Phys. Rev. E 101, 042304 (2020)]\cite{newmanimproved} proposed reduced mutual information (RMI), trying to handle the problems of NMI. This comment discusses why the examples used in the article are not accurate, and empirically demonstrates that RMI cannot overcome the challenges and it even brings more.

Firstly, Consider a set of $n$ objects with ground-truth division $\pi_1$. $\pi_1$ is nontrivial, i.e., the number of clusters $c$ is larger than 1. Suppose that there are two divisions $\pi_2$ and $\pi_3$ obtained by some clustering methods,
where $\pi_2$ consists of just a single cluster containing all the objects and $\pi_3$ consists of $n$ clusters, each containing a single object. Actually, $\pi_2$ and $\pi_3$ are not unrelated with $\pi_1$. For example, if $\pi_1$ is $[1,\, 1,\, 1,\, 1,\, 1,\, 1,\, 2,\, 2,\, 3,\, 3]$, there are at least two objects clustered correctly in $\pi_2$ and three objects clustered correctly in $\pi_3$, meaning that the value of any index between $\pi_1$ and $\pi_2$ or between $\pi_1$ and $\pi_3$ may not be zero. Indeed, the value of any index between $\pi_1$ and $\pi_3$ should be larger than that between $\pi_1$ and $\pi_2$, because there is only shared information in one single cluster between $\pi_1$ and $\pi_2$  while there is shared information in $c$ clusters between $\pi_1$ and $\pi_3$. In other words, $\pi_2$ only (partly) reveals information of a single cluster, but $\pi_3$ (partly) reveals information of $c$ clusters, meaning that $\pi_3$ is more informative. Generally speaking,  the index value between two random and independent divisions depends on the objects size $n$ and the cluster number $c$, and may not be zero or some constant. In extreme circumstances where $c=n$, the reasonable value between $\pi_1$ and $\pi_3$ should be 1. RMI does not act in this way, actually, $\mbox{RMI}(\pi_3,\,\pi_3) = 0$. The authors explains this result from the perspective of information theory.

The problem of clustering evaluation is in principle a kind of comparison problem, not a kind of problem in information theory. We can use theory of information encoding for comparison between two divisions. However, if the results do not meet intuitions, we need to use other methods instead of information encoding. An intuition is that if some method reveals the ground-truth division perfectly on a dataset, the method would also reveal the ground-truth division perfectly on its subset. In other words, any measure between the ground-truth division on a dataset against itself should be 1, and the measure between the ground-truth division on subset against itself should also be 1. However, RMI does not.

Now, imagine that there are 300 objects $\{a_i:\,i=1,\,2,\,\cdots,\,300\}$  with ground-truth division $\pi_1$:
\begin{equation*}
\begin{array}{cccccc}
[1   & 1   & ... & ... &  1   &  1   \\
 2   & 2   & ... & ... &  2   &  2   \\
 ... & ... & ... & ... & ...  & ...  \\
 ... & ... & ... & ... & ...  & ...  \\
 30  & 30  & ... & ... & 30   & 30]   \\
\end{array}
\end{equation*}

The number of clusters is 30. If the division $\pi$ obtained by some method is identical with $\pi_1$, the index value $\mbox{RMI}(\pi_1, \pi) $ is obviously a large positive number. Now consider the subset: $\{a_i:\,i=10,\,20,\,\cdots,\,300\}$. Again, the method is applied on it, but this time the value of RMI is 0. Similarly, consider the subset:  $\{a_i:\,i=1,\,2,\,3,\,\cdots,\,10\}$, the value of RMI is still 0. In summary, the performance of the method depends on which dataset we used. Its performance is perfect on the complete set, but poor on a subset. This is hard to explain.

The authors may argue that the result of zero does not necessarily mean a poor result. If this is true, there are two kinds of zeros introduced by RMI: one means a poor result and the other means a perfect result. This will be very misleading.

If $\mbox{RMI}(\pi_3,\,\pi_3) = 0$ is reasonable, for the data containing   $n$ objects, and the ground-truth division $\pi_3$, we cannot design any clustering methods analyzing it actually, because the result of any method is zero.

In summary, the key problem here is how to explain $\mbox{RMI}(\pi_3,\,\pi_3) = 0$. If $\mbox{RMI}(\pi_3,\,\pi_3) = 0$  means a poor result, there is no hope to analyze the data with $n$  clusters since no method can output good result on it, and it is possible that performance of some method is perfect on the complete set, but poor on a subset. Otherwise, $\mbox{RMI}(\pi_3,\,\pi_3) = 0$  does not mean a poor result, then there are two kinds of zeros. Hence $\mbox{RMI}(\pi_3,\,\pi_3) = 0$  is a source of much confusion, and one needs to be aware of this. To the best of our knowledge,  RMI is the only measure that gives zero value of a division against itself.

\begin{figure*}
  \includegraphics[height=70mm,width=85mm]{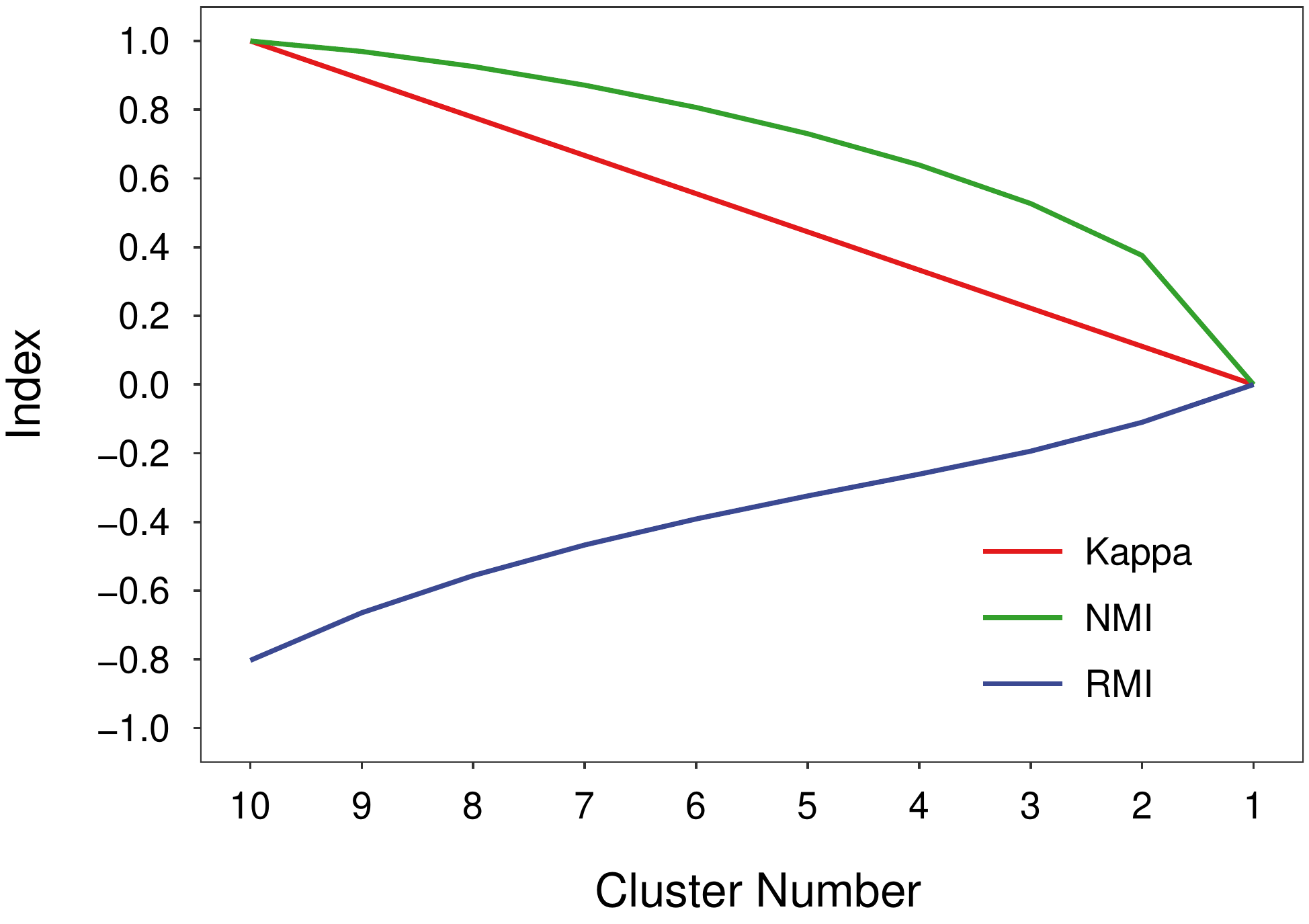}
  \caption{Kappa, NMI and RMI index of two divisions $\pi_3$ and $\pi'$. $\pi_3$ consists of 10 clusters, each containing a single object. $\pi'$ is obtained through merging clusters in $\pi_3$ gradually until there is only one single cluster left containing all the objects.}\label{Fig:01}
\end{figure*}

We calculates the values of NMI, RMI\footnote{The codes are available from the homepage of Prof. Newman: http://www-personal.umich.edu
/~mejn/. Although in README, it said: ``There can be any number of objects and any number of groups.'', the codes are more valid in the case where the number of clusters $c$ is substantially smaller than the number of objects $n$ since they use the approximation Eq.(29). We use the codes for the comparison anyway.} and Kappa \cite{liu2019evaluation} between $\pi_3$ and $\pi'$, where $\pi'$ is obtained through merging clusters in $\pi_3$ gradually until there is only one single cluster left containing all the objects. The results are given in Fig.\ref{Fig:01}, from which, one can observe that: 1) RMI increases from negative value to zero. The negative values, especially for $\pi'$ starting to deviate from $\pi_3$, are  hard to explain. Note that this point may not be true since the codes are not suitable in this case. Please see next paragraph for more details. 2) The line of NMI is not proportionally decreased. 3) The line of Kappa is straight, satisfying proportionality assumption.

Secondly, as the article has pointed out, Eq. (23) only works for small $n$  and $c$. One has to approximate it using Eq. (28) or Eq. (29). In the article, the authors said, ``We are particularly interested in two limits. The first is the sparse limit, typified by our example above in which each object is placed in a group on its own. $\cdots$ Usually we are interested in cases where the numbers of groups R and S are substantially smaller than $n$.''  The above sentences can be understood in two ways: 1) the first case is that the number of clusters $c$ is $n$, and the other case is that the number of clusters $c$ is not $n$. If this is true, the codes are used correctly in Fig. \ref{Fig:01} when $\pi'\neq\pi$, and the results are counter-intuitive. 2) The first case is that the number of clusters  $c$ is large and is compatible with $n$, and the other case is that the number of clusters $c$   is much smaller than $n$. Consequently, our discussion is based on this understanding.

The value of $\mbox{RMI}(\pi_3,\,\pi_3)$ is -0.8 using the approximation Eq.(29) and is 0 without approximation, meaning that different approximations may lead to very different results. Now, consider a set of $n$  objects with ground-truth division $\pi_1$. The number of clusters $c$ in $\pi_1$ is substantially smaller than  $n$. There are two divisions $\pi_2$  and $\pi_3$  obtained by some methods, which belong to different cases, i.e., the number of clusters in $\pi_2$  is large and that in $\pi_3$  is small. Then we need to use different approximations for the calculation of $\mbox{RMI}(\pi_1,\,\pi_2)$  and $\mbox{RMI}(\pi_1,\,\pi_3)$ . Is this comparison fair since different approximations may lead to very different results? In other words,  $\mbox{RMI}(\pi_1, \pi_3) > \mbox{RMI}(\pi_1, \pi_2)$ may be less about the goodness of  $\pi_3$, but more about the approximation.

Furthermore, we calculate RMI of $\pi_4$ against itself, where $\pi_4$  is a division on 2000 objects and the number of clusters  $c$ is 200. We believe that Eq.(29) works for this case, and the codes provided by the authors can be used. The result is:  $\mbox{RMI}(\pi_4, \pi_4) =-3.45$, which is still counter intuitive.

In summary, there is not a clear boundary between the two cases, and there is no discussion on the relations between the approximations and the comparison of different results, making it hard to choose the suitable approximation. Counter-intuitive phenomenon still exists even if we use appropriate approximation.

Thirdly, the title is not accurate, because MI or NMI is only used for clustering and community detection evaluation. The main challenge here is the limited amount of information available from the division result. The labels in the division can only tell us which objects are clustered together and which ones are not. For example, the two divisions $[1,\, 1,\, 1,\, 1,\, 1,\, 1,\, 2,\, 2,\, 3,\, 3]$ and $[2,\, 2,\, 2,\, 2,\, 2,\, 2,\, 3,\, 3,\, 1,\, 1]$ are actually identical. One has to define specific criteria including Rand index and NMI, ``which are invariant under permutations of the labels'', as the authors said. The above problems can be largely mitigated in classification since the computed labels are relatively more informative, making point-wise label comparison possible and reasonable. One can use more powerful criteria for classification evaluation such as Kappa. In our previous work \cite{liu2019evaluation}, clustering and classification evaluations are connected through linear programming and Kappa is employed for clustering evaluation.

Finally, we'd like to summarize the problems with NMI and its variants: 1) ignoring importance of small clusters. Table \ref{Tab:01} gives the results of several indices between $\pi_1$ and $\pi_5$, and $\pi_1$ and $\pi_6$, where $\pi_{1} = [1,\, 1,\, 1,\, 1,\, 1,\, 1,\, 2,\, 2,\, 3,\, 3]$, $\pi_{5} = [1,\, 1,\, 1,\, 1,\, 1,\, 2,\, 2,\, 2,\, 3,\, 3]$ and
$\pi_{6} = [1,\, 1,\, 1,\, 1,\, 1,\, 1,\, 2,\, 2,\, 2,\, 3]$. Values of Kappa are more reasonable than NMI and its variants. The table also gives the indices between $\pi_i$ against itself, $i = 1,\,2,\,3$, where $\pi_{2} = [1,\, 1,\, 1,\, 1,\, 1,\, 1,\, 1,\, 1,\, 1,\, 1]$ and
$\pi_{3} = [1,\, 2,\, 3,\, 4,\, 5,\, 6,\, 7,\, 8,\, 9,\, 10]$. The values of RMI against itself are not 1, bringing new difficulty for clustering evaluation.
   2) violating the so-called proportionality assumption (Fig. \ref{Fig:01}). 3) being not able to evaluate specific clusters.

 In short, the standard NMI is not perfect. The proposed RMI does not handle the issues and even brings new difficulty. The necessity of Kappa is empirically validated in this comment.

\begin{table}[H]
\renewcommand\arraystretch{1.5}
\centering
\tabcolsep 15pt 
\caption{Summary of several indices between two divisions $a$ and $b$: $\left<a, b\right>$. The values of RMI between $a$ against itself are not 1, bringing new difficulty for clustering evaluation. For RMI, the values outside the brackets are obtained by Eq.(23), and the ones inside the brackets are by Eq.(29), meaning that different approximations may lead to very different results.}
\begin{tabular}{cccccc}
\toprule\hline\hline
  &$\left<\pi_{1}, \pi_{5}\right>$ & $\left<\pi_{1}, \pi_{6}\right>$ &$\left<\pi_{1}, \pi_{1}\right>$ & $\left<\pi_{2}, \pi_{2}\right>$  &$\left<\pi_{3}, \pi_{3}\right>$ \\
\hline
NMI\cite{danon2005comparing} & 0.77 & 0.82 & 1 & 1 & 1 \\
AMI\cite{vinh2010information} & 0.64 & 0.73 & 1 & 1 & 1 \\
ARI\cite{vinh2010information} & 0.66 & 0.86 & 1 & 1 & 1 \\
V-measure\cite{rosenberg2007v} & 0.77 & 0.82 & 1 & 1 & 1 \\
VI\cite{meilua2007comparing} & 0.67 & 0.48 & 0 & 0 & 0 \\
Q$_2$\cite{dom2002information} & 0.70 & 0.73 & 1 & 1 & 1 \\
RMI\cite{newmanimproved} & \ul{1.27 (0.38)} & \ul{1.46 (0.44)} & \ul{1 (0.57)} & \ul{0 (0)} & \ul{0 (-0.80)} \\
Kappa\cite{liu2019evaluation} & 0.83 & 0.82 & 1 & 1 & 1 \\
\hline\hline
\bottomrule
\end{tabular}
\label{Tab:01}
\end{table}

\bibliography{ref}

\end{document}